\documentclass{PoS}
\usepackage{amssymb}
\usepackage{amsmath}
\usepackage[]{bbm}
\newcommand{\hZ}{\hat{Z}}
\newcommand{\hM}{\hat{M}}
\newcommand{\be}{\begin{equation}}
\newcommand{\ee}{\end{equation}}
\newcommand{\Tr}{\textrm{Tr}}
\newcommand{\vc}[1]{\mbox{\boldmath $#1$}}
\newcommand{\dd}{\textrm{d}}

\newcommand{\hlf}{\frac{1}{2}}

\newcommand{\myRe}{\textrm{Re}}
\newcommand{\MSb}{\overline{\rm{MS}}}

\newcommand{\Det}{\textrm{Det}}

\title{Z(3)-symmetric effective theory of hot QCD}

\ShortTitle{Z(3)-symmetric effective theory of hot QCD}

\author{\speaker{Aleksi Kurkela}%
         \\
        Department of Physical Sciences P.O.Box 64 FI-00014 University of Helsinki, Finland\\
        E-mail: \email{aleksi.kurkela@helsinki.fi}}

\abstract{
We study a three dimensional Z(3)-symmetric effective theory of high temperature QCD. The exact lattice-continuum relations, needed in order to perform lattice simulations with physical parameters, are computed
to order $\mathcal{O}(a^0)$ in lattice perturbation theory.
Lattice simulations are performed to determine the phase structure of a subset of the parameter space.}

\FullConference{The XXV International Symposium on Lattice Field Theory\\
		 July 30 - August 4 2007\\
		 Regensburg, Germany}

\begin{document}

\section{Introduction}
At high temperature, QCD matter undergoes a deconfinement transition, where ordinary had\-ron\-ic matter transforms into strongly interacting quark-gluon plasma. In the absence of quarks, $N_f=0$, the transition is a symmetry-breaking first order transition, where the order parameter is the thermal Wilson line. The non-zero expectation value of the Wilson line signals the breaking of the Z(3) center symmetry of quarkless QCD at high temperatures.

The transition has been studied extensively using lattice simulations \cite{Blum:1994zf}, but becomes computationally exceedingly expensive at high temperatures $T\gtrsim 5 T_c$. At high $T$, the complementary approach has been to construct perturbatively effective theories, such as EQCD and MQCD, using the method of dimensional reduction \cite{Ginsparg:1980ef}.  In the dimensional reduction procedure, however, one expands the temporal gauge fields around one of the Z(3) vacua and thus explicitly violates the center symmetry and the effective theories fail to describe QCD for $T$ below $5T_c$.

As a unification of these strategies, a 3D effective theory of hot QCD respecting the Z(3) symmetry has been constructed in \cite{Vuorinen:2006nz}. At high temperatures, the effective theory is matched to EQCD but still preserves the center symmetry. The effective theory is further connected to full QCD by matching the domain wall profile separating two different Z(3) minima. Thus, one hopes that the range of validity of this theory would extend down to $T_c$.

In order to perform lattice simulations the theory has been formulated on a lattice and the lattice theory has been matched to the continuum theory in \cite{Kurkela:2007dh}. The effective theory is super-renormalizable, and thus the exact relations between the continuum $\MSb$ and lattice regulated theories can be obtained via two-loop lattice perturbation theory.

\section{Theory}
The theory we are studying is defined by a three dimensional continuum action, which we renormalize in the $\MSb$ scheme
\begin{equation}
S = \int \dd ^{3-2\epsilon}x  \left\{ \hlf \Tr F_{ij} ^2  + \Tr\left( D_i Z^\dagger D_i Z \right) + V_0(Z) + V_1(Z) \right\},\label{action}
\end{equation}
where
\begin{align}
F_{ij} &=\partial_i A_j - \partial_j A_i + i g_3 [A_i,A_j], \quad \quad D_i = \partial_i - i g_3 [A_i,\quad]
\end{align}
and $Z$ is a $3\times3$ complex matrix, which in the limit $\epsilon\rightarrow 0$ has dimension $\dim Z = \sqrt{\textrm{GeV}}$. The gauge fields $A_i$ are Hermitean traceless $3\times 3$ matrices and can be expressed using generators of SU(3), $A_i=A_i^a T^a$, with $\Tr T^aT^b=\hlf\delta^{ab}$ . The covariant derivative is in the adjoint representation. The potentials are
\begin{align}
V_0(Z) &= c_1 \Tr[Z^\dagger Z] +2c_2 \myRe(\Det[Z])+c_3 \Tr[(Z^\dagger Z)^2], \label{V0}\\
V_1(Z) &=  d_1 \Tr[M^\dagger M] + 2 d_2 \myRe(\Tr[M^3]) + d_3 \Tr[(M^\dagger M)^2],\label{V1}
\end{align}
where $M=Z-\frac{1}{3}\Tr[Z] \mathbbm{1}$ is the traceless part of $Z$. Here, the gauge coupling $g_3$ has a positive mass dimension $\dim[g_3^2] =$GeV, making the theory
super-renormalizable. Because of the super-renormalizability, the coefficients $c_2,c_3,d_2$, and $d_3$ are renormalization scale independent and only the mass terms $c_1$ and $d_1$ acquire a scale dependence in the $\MSb$ renormalization scheme. 
The coefficients $c_i$, $d_i$, and $g_3$ can be matched to the parameters of full thermal QCD by imposing the conditions that the theory reduce to EQCD at the high temperature limit, and that the theory reproduce the correct domain wall profile of full QCD \cite{Vuorinen:2006nz}. This defines a subset of parameter values (with a limited accuracy due to perturbative matching), for which the theory describes thermal QCD. However, we consider here the theory in general, and do not restrict ourselves only to the perturbative matching regime.

The action is invariant under local gauge transformations, with $Z$ transforming in the adjoint representation:
\begin{align}
A_i(\vc{x}) &\longrightarrow G(\vc{x}) \left(A_i(\vc{x}) -\frac{i}{g_3}\partial_i \right)G^{-1}(\vc{x}), \\
Z(\vc{x})  &\longrightarrow G(\vc{x}) Z(\vc{x}) G^{-1} (\vc{x}),
\end{align}
where $G(\vc{x})\in$SU(3). In addition to the local transformations, the action is invariant under the global Z(3) transformation
\begin{equation}
Z\longrightarrow e^{i2\pi n/3} Z, \quad n=1,2,\ldots
\end{equation}

\section{Lattice action}
The lattice action corresponding to the continuum theory can be written as $S = S_W + S_Z$, 
where 
\begin{align}
S_W = \beta &\sum_{x,i<j} \left[ 1 - \frac{1}{3} \myRe \Tr[ U_{\mu\nu}] \right] 
\end{align}
is the standard the Wilson action with the lattice coupling constant $\beta = 6/(a g_3^2)$.

The kinetic term, $\Tr\left( D_i Z^\dagger D_i Z \right)$, is discretized by replacing the covariant derivatives by covariant lattice differences. Then the scalar sector of the action reads: 
\begin{align}
 S_Z = 2\left(\frac{  6 }{ \beta } \right)& \sum_{x,i}\myRe \Tr \left[\hZ^\dagger\hZ - \hZ^\dagger(x)U_i(x)\hZ(x+\hat{i})U^\dagger_i(x)\right] \\
+ \left(\frac{6 }{ \beta}\right) ^3 & \sum_x \left( \hat{c}_1\Tr[\hZ^\dagger \hZ] + 2\hat{c}_2 \myRe\Det{\hZ} + \hat{c}_3 \Tr[(\hZ^\dagger \hZ)^2] + \hat{d}_1\Tr[\hM^\dagger \hM] + 2\hat{d}_2 \myRe\Tr{\hM^3} + \hat{d}_3 \Tr[(\hM^\dagger \hM)^2] \right)\nonumber.
\end{align}
where $\hat{c}_i, \hat{d}_i,\hM$, and $\hZ$ are dimensionless. Only the mass terms $\hat{c}_1$ and $\hat{d}_1$ require non-trivial renormalization and 
all the other terms can be matched to order $\mathcal{O}(a^0)$ on tree-level by simply scaling with $g_3$:
\begin{eqnarray}
Z  =g_3 \hZ,         & M   = g_3 \hM \\
c_2=g_3^3 \hat{c}_2, & d_2 = g_3^3 \hat{d}_2\\
c_3=g_3^2 \hat{c}_3,       & d_3 = g_3^2\hat{d}_3.
\end{eqnarray}
For the mass terms, renormalization has to be carried out, so that the long distance physics is the same in both regularization schemes. A two-loop lattice perturbation theory calculation gives:
\begin{align}
\hat{c}_1 =& \frac{c_1}{g_3^4} - \frac{1}{4 \pi} 6.3518228 \hat{c}_3 \beta
  -\frac{1}{16\pi^2}\left[\left(64 \hat{c}_3 + \frac{88}{9} \hat{c}_3^2 \right) \left( \log \beta + 0.08849 \right)  + 37.0863  \hat{c}_3 \right]  + \mathcal{O}(\beta^{-1}) 
\label{c1_ren}
\end{align}
and
\begin{align}
\hat{d}_1 =& \frac{d_1}{g_3^4} - \frac{\beta}{4 \pi} \left(3.17591 + 5.64606  \hat{d}_3 \right)
 - \frac{1}{16 \pi^2} \Big\{41.780852 +37.0863 \hat{d}_3 \nonumber\\
-& \left( \frac{280}{9} \hat{c}_3^2 - 64 \hat{d}_3 + \frac{184}{3}\hat{d}_3 \hat{c}_3 + \frac{92}{3}\hat{d}_3^2 + \frac{9}{2}\right) \left[\log\beta + 0.08849\right] \Big\}+\mathcal{O}(\beta^{-1}). \label{d1_ren}
\end{align}
There are also higher order corrections (corrections of order $\mathcal{O}(\beta^{-1})$ corresponding to order $\mathcal{O}(a)$ in lattice spacing), but their effect vanishes in the continuum limit. 
Various operators have also been renormalized in \cite{Kurkela:2007dh} on the lattice in order to convert their expectation values to continuum regularization.

\section{Phase diagram of $V_1(Z)$}
A simpler model is obtained from the original theory by setting $c_i=0$. In this model, the trace of $Z$ decouples and can be integrated over as a free scalar field and the relevant degree of freedom is thus a traceless complex matrix $M$.
The model is defined by the action:
\begin{align}
S &= \int \dd^3x \left[\frac{1}{2}\Tr F_{ij}^2+ \Tr D_i M^\dagger D_i M + d_1 \Tr M^\dagger M+2d_2 \myRe(\Tr[M^3])+d_3 \Tr( M^\dagger M)^2\right]. 
\end{align}
If the cubic term $d_2$ is zero, the Lagrangian is invariant under a U(1) global symmetry $M \rightarrow gM$, $g \in $U(1). The breaking of the symmetry is signalled by a local order parameter:
\be
	\mathcal{A}=\sqrt{\langle\Tr (M+M^\dagger)^3\rangle^2 + \langle\Tr ( M-M^\dagger)^3\rangle^2}.
\ee
In the symmetric phase $\mathcal{A}$ is strictly zero and in the broken phase it has a non-zero vacuum expectation value, while the two phases are separated by a first order transition. In the broken phase $\langle \Tr M^\dagger M\rangle$ is larger than in the symmetric phase.
After the inclusion of the cubic term, $\mathcal{A}$ is no longer strictly an order parameter, since the U(1) symmetry is explicitly broken. However, the first order transition remains and is accompanied with a significant discontinuity in $\mathcal{A}$ and $\langle \Tr M^\dagger M \rangle$.

\subsection{Lattice analysis}
A non-perturbative lattice analysis has been performed to obtain the phase structure of the model.  For the simulations we used a hybrid Monte-Carlo algorithm for the scalar fields and Kennedy-Pendleton quasi heat bath and full group overrelaxation for the link variables.

\begin{figure}[t!]
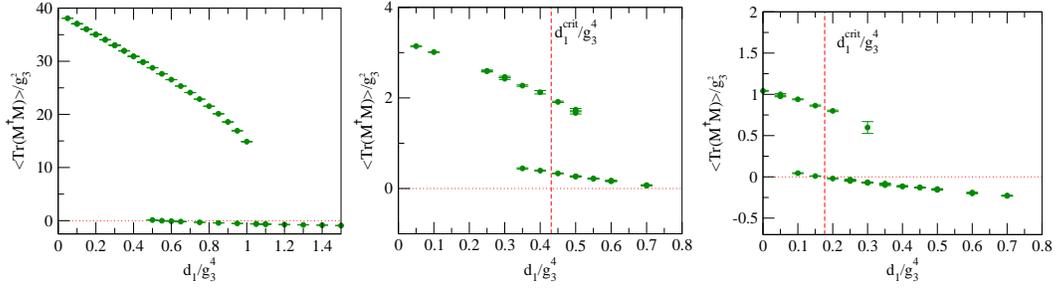

\begin{center}
\includegraphics*[width = 0.3\textwidth]{figure8a.eps}
\includegraphics*[width = 0.3\textwidth]{figure8b.eps}
\includegraphics*[width = 0.31\textwidth]{figure8c.eps}
\caption{Discontinuity in the quadratic condensate in continuum regularization $\langle\Tr M^\dagger M\rangle_{\MSb}$ for $d_3=0.1,1,3$. The phase transition gets weaker as the coupling $d_3$ grows. The metastable regions shrink and the discontinuity diminishes.}\label{hyster}\end{center}
\end{figure}

\begin{figure}[ht]
\begin{center}
\includegraphics*[width = 0.6\textwidth]{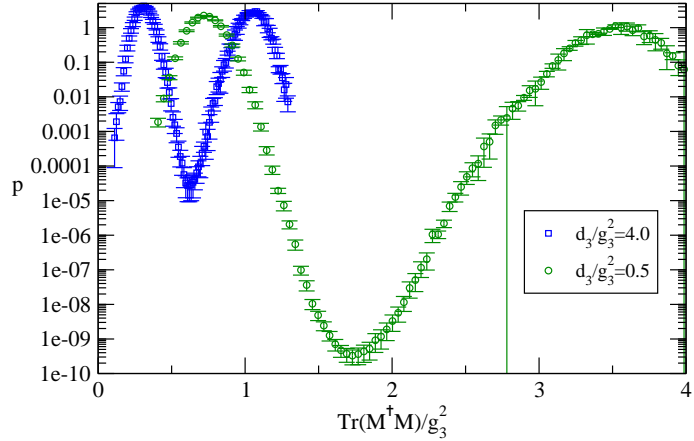}
\caption{Histograms of $\Tr M^\dagger M$ in logarithmic scale with $d_2=0$ along the critical curve. Transition channel between the peaks weakens and the transition gets stronger for decreasing $d_3$. For $d_3/g_3^2=0.5$, the relative probability density in the tunneling channel is suppressed by a factor $\sim10^{-10}$.}\label{histgram}\end{center}
\end{figure}

The transition was found to be of the first order for all parameter values used in the simulations  ($d_3\leq4$ and $d_2\leq0.15$) accompanied with a large latent heat and surface tension; hysteresis curves showing discontinuity around critical point in $\langle \Tr M^\dagger M \rangle_{\MSb}$ can be seen in Fig.\ref{hyster}. The probability distributions of $\Tr M^\dagger M$ along the critical curve are very strongly separated (see Fig.\ref{histgram}). This makes the system change its phase very infrequently during a simulation, and a multicanonical algorithm is needed to accommodate a phase flip in reasonable times for any system of a modest size. Even with the multicanonical algorithm, the critical slowing restricts us to physical volumes up to $V\lesssim 50/g_3^6$. 

\begin{figure}[ht]
\begin{center}
\includegraphics*[width = 8.6cm]{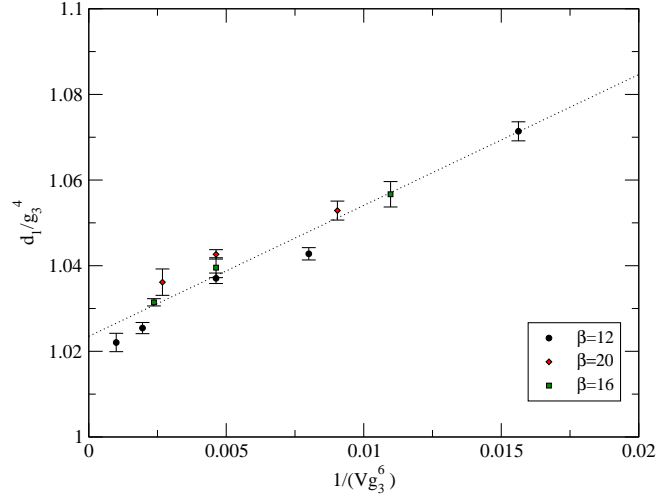}
\caption{Volume and lattice spacing dependence of the pseudo-critical point with $d_3=2$ and $d_2=0.1$. The pseudo-critical point was determined by requiring equal probability weight for $\Tr M^\dagger M$ in both phases. The lines represent linear fits. The dependence on lattice spacing and volume seem to be within 5\% for the lattice spacings and volumes used. }\label{finite_size}\end{center}
\end{figure}

The pseudo-critical point was determined requiring equal probability weight for $\Tr M^\dagger M$ in both phases.  The simulations were performed with $\beta=12$ and a lattice size $N^3=12^3$, which precludes the continuum extrapolation as well as the thermodynamical limit. However, these limits were studied for one set of parameter values. The dependence of the critical point on the lattice spacing was beyond our resolution for the lattice spacings used and the volume dependence was found to be of order of five per cent for the volumes used (see Fig.\ref{finite_size} ).

\begin{figure}[hb!]
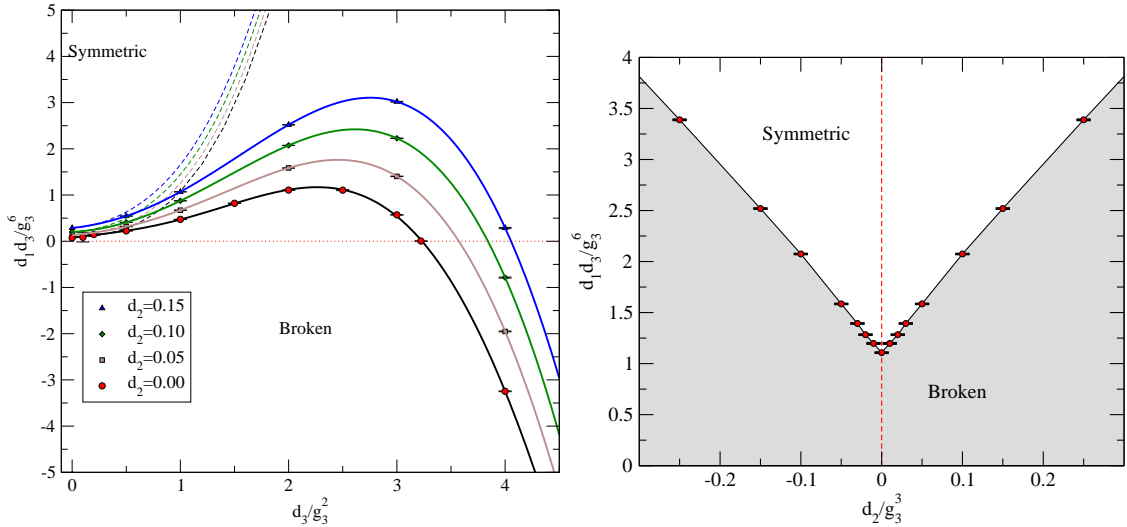

\begin{center}
\includegraphics*[width = 0.49\textwidth]{figure9.eps}
\includegraphics*[width = 0.49\textwidth]{figure10.eps}
\caption{The phase diagram of the soft potential as a function of $d_1,d_2$ and $d_3$. First order critical line separates two phases. Solid lines represent polynomial fits to the lattice data points and dashed lines are the perturbative predictions. The symmetric phase refers to the phase where with $d_2=0$ the order parameter vanishes and with $d_2\neq0$ is smaller than in the broken phase. On the right panel $d_3=2$.}\label{phasediagram5}\end{center}
\end{figure}

The phase diagram can be seen in Fig.\ref{phasediagram5}. The non-perturbative critical line measured from the lattice follows the one-loop perturbative result for small values of $d_3$, but for larger $d_3$ fluctuations make the system prefer the symmetric phase. The discontinuity in $\langle\Tr M^\dagger M\rangle$ along the critical line diminishes, as $d_3$ gets larger, but it seems that the discontinuity persists, even if its magnitude diminishes in the limit $d_3\rightarrow\infty$ suggesting that there is a first order phase transition for any (positive) value of $d_3$.

\section{Conclusions}
The exact relations between the lattice and continuum $\MSb$ regulated formulations of the Z(3)-symmetric 3D effective theory of hot QCD have been calculated in \cite{Kurkela:2007dh}. The Lagrangians and the operators up to cubic ones have been matched to $\mathcal{O}(a^0)$. These results make the non-perturbative lattice study of the theory possible. 

An interesting model  with non-trivial dynamics is obtained by setting $c_i=0$ in Eq.(\ref{V0}). The phase diagram of this model has been determined using lattice simulations. Two distinct phases were found, separated by a strong first order transition.

In the future, it is our goal to map out the phase 
diagram in the full parameter space of the theory, in order to search for regions in which
the phase diagram would resemble that expected for 
the finite-temperature SU(3) pure Yang-Mills theory. 

\section*{Acknowledgments}
This research has been supported by Academy of
Finland, contract number 109720 and the EU I3 Activity
RII3-CT-2004-506078 HadronPhysics. Simulations were carried out at CSC - Scientific Computing Ltd., Finland


\begin{thebibliography}{99}
\bibitem{Blum:1994zf}
T.~Blum, L.~K{\"a}rkk{\"a}inen, D.~Toussaint, and S.~A. Gottlieb, {\it The beta
  function and equation of state for {QCD} with two flavors of quarks},  {\em
  Phys. Rev.} {\bf D51} (1995) 5153--5164,
  [\href{http://xxx.lanl.gov/abs/hep-lat/9410014}{{\tt hep-lat/9410014}}];\\
F.~Karsch, E.~Laermann, and A.~Peikert, {\it The pressure in 2, 2+1 and 3
  flavour {QCD}},  {\em Phys. Lett.} {\bf B478} (2000) 447--455,
  [\href{http://xxx.lanl.gov/abs/hep-lat/0002003}{{\tt hep-lat/0002003}}];\\
U.~M. Heller, {\it Recent progress in finite temperature lattice {QCD}},  {\em
  PoS} {\bf LAT2006} (2006) 011,
  [\href{http://xxx.lanl.gov/abs/hep-lat/0610114}{{\tt hep-lat/0610114}}].

\bibitem{Ginsparg:1980ef}
P.~Ginsparg, {\it First order and second order phase transitions in gauge
  theories at finite temperature},  {\em Nucl. Phys.} {\bf B170} (1980) 388;\\
T.~Appelquist and R.~D. Pisarski, {\it Hot {Y}ang-{M}ills theories and
  three-dimensional {QCD}},  {\em Phys. Rev.} {\bf D23} (1981) 2305;\\
K.~Kajantie, M.~Laine, K.~Rummukainen, and M.~E. Shaposhnikov, {\it Generic
  rules for high temperature dimensional reduction and their application to the
  standard model},  {\em Nucl. Phys.} {\bf B458} (1996) 90--136,
  [\href{http://xxx.lanl.gov/abs/hep-ph/9508379}{{\tt hep-ph/9508379}}];\\
E.~Braaten and A.~Nieto, {\it Free energy of {QCD} at high temperature},  {\em
  Phys. Rev.} {\bf D53} (1996) 3421--3437,
  [\href{http://xxx.lanl.gov/abs/hep-ph/9510408}{{\tt hep-ph/9510408}}];\\
K.~Kajantie, M.~Laine, K.~Rummukainen, and Y.~Schr{\"o}der, {\it The pressure
  of hot {QCD} up to {g**6 ln(1/g)}},  {\em Phys. Rev.} {\bf D67} (2003)
  105008, [\href{http://xxx.lanl.gov/abs/hep-ph/0211321}{{\tt
  hep-ph/0211321}}];\\
A.~Hietanen and A.~Kurkela, {\it Plaquette expectation value and lattice free
  energy of three-dimensional {SU(N)} gauge theory},  {\em JHEP} {\bf 11}
  (2006) 060, [\href{http://xxx.lanl.gov/abs/hep-lat/0609015}{{\tt
  hep-lat/0609015}}];\\
M.~Veps{\"a}l{\"a}inen, {\it Mesonic screening masses at high temperature and
  finite density},  {\em JHEP} {\bf 03} (2007) 022,
  [\href{http://xxx.lanl.gov/abs/hep-ph/0701250}{{\tt hep-ph/0701250}}].




 \bibitem{Vuorinen:2006nz}
  A.~Vuorinen and L.~G.~Yaffe,
  {\it Z(3)-symmetric effective theory for SU(3) Yang-Mills theory at high
  temperature,}
  Phys.\ Rev.\  D {\bf 74} (2006) 025011,
 [\href{http://xxx.lanl.gov/abs/hep-ph/0604100}{{\tt hep-ph/0604100}}].

\bibitem{Kurkela:2007dh}
  A.~Kurkela,
  {\it Framework for non-perturbative analysis of a Z(3)-symmetric effective
  theory of finite temperature QCD,}
  Phys.\ Rev.\ D {\bf 76} (2007) 094507, 
 [\href{http://xxx.lanl.gov/abs/0704.1416}{{\tt arXiv:0704.1416}}].
\end{thebibliography}
\end{document}